\documentclass{aa}
\usepackage[varg]{txfonts}
\usepackage{graphicx}
\usepackage{natbib}
\usepackage{hyperref}
\usepackage{color}
\bibpunct{(}{)}{;}{a}{}{,} 

\newcommand{\blue}[1]{{\color{blue}{#1}}}

\newcommand{\acrit}{$a_{crit}$}

\newcommand{\amax}{$a_{crit, max}$ }
\newcommand{\aratio}{$a_{crit, max}/a_{crit, init}$ }
\newcommand{\acbp}{$a_{cbp}$ }
\newcommand{\ecbp}{$e_{cbp}$ }
\newcommand{\efree}{$e_{free}$ }
\newcommand{\abin}{$a_{bin}$ }
\newcommand{\ebin}{$e_{bin}$ }

\newcommand{\Msun}{$M_{\odot}$}
\newcommand{\rg}{$r_g$}
\newcommand{\vplanet}[0]{\texttt{VPLanet}\xspace}
\newcommand{\eqtide}[0]{\texttt{EqTide}\xspace}
\newcommand{\stellar}[0]{\texttt{STELLAR}\xspace}
\newcommand{\binary}[0]{\texttt{BINARY}\xspace}
\newcommand{\kepler}[0]{{\it Kepler}\xspace}
\newcommand{\link}[1]{{\small{\texttt{\underline{#1}}}}}

\def\eg{{\it e.g.,}\xspace}

\def\cf{{\it c.f.,}\xspace}
\def\gsim{~\rlap{$>$}{\lower 1.0ex\hbox{$\sim$}}}
\def\lsim{~\rlap{$<$}{\lower 1.0ex\hbox{$\sim$}}}

\begin{document}
\title{Orbital evolution of potentially habitable planets of tidally interacting binary stars}
\author{David E. Graham\inst{1,2}
  \and David P. Fleming\inst{2,3}
  \and Rory Barnes\inst{2,3}}

\institute{Institut für Theoretische Astrophysik (ITA), Zentrum für Astronomie (ZAH), Universität Heidelberg, Albert-Ueberle-Str.2 69120 Heidelberg, Germany
  \and Astronomy Department, University of Washington, Box 951580, Seattle, WA, 98195, USA
  \and NASA Virtual Planetary Laboratory, Seattle, WA, 98195, USA}

\date{Received 16 July 2020 / Accepted 2 November 2020}

\abstract{We simulate the coupled stellar and tidal evolution of short-period binary stars (orbital period $P_{orb} \lsim$8 days) to investigate the orbital oscillations, instellation cycles, and orbital stability of circumbinary planets (CBPs). We consider two tidal models and show that both predict an outward-then-inward evolution of the binary's semi-major axis $a_{bin}$ and eccentricity $e_{bin}$. This orbital evolution drives a similar evolution of the minimum CBP semi-major axis for orbital stability. By expanding on previous models to include the evolution of the mass concentration, we show that the maximum in the CBP orbital stability limit tends to occur 100 Myr after the planets form, a factor of 100 longer than previous investigations. This result provides further support for the hypothesis that the early stellar-tidal evolution of binary stars has removed CBPs from short-period binaries. We then apply the models to Kepler-47 b, a CBP orbiting close to its host stars' stability limit, to show that if the binary's initial $e_{bin} \gsim$0.24, the planet would have been orbiting within the instability zone in the past and probably wouldn't have survived. For stable, hypothetical cases in which the stability limit does not reach a planet's orbit, we find that the amplitudes of $a_{bin}$ and $e_{bin}$ oscillations can damp by up to 10\% and 50\%, respectively. Finally, we consider equal-mass stars with $P_{orb} =$ 7.5 days and compare the HZ to the stability limit. We find that for stellar masses $\lsim0.12M_{\odot}$, the HZ is completely unstable, even if the binary orbit is circular. For $e_{bin} \lsim$0.5, that limit increases to $0.17M_{\odot}$, and the HZ is partially destabilized for stellar masses up to $0.45M_{\odot}$. These results may help guide searches for potentially habitable CBPs, as well as characterize their evolution and likelihood to support life after they are found.} 

  \maketitle

\section{Introduction}

The discovery of circumbinary planets (CBPs) has opened up the possibility that we may discover some that support liquid surface water and detectable biosignatures. A key driver for sustaining liquid water on Earth's surface is a nearly stable solar radiation flux that provides enough energy to melt ice, but not so much as to boil water. The incident stellar radiation (instellation) on a CBP necessarily evolves due to the time-varying gravitational interactions of its two host stars, potentially affecting the long-term stability of liquid water on CBPs. While this dynamical feature of CBPs is well known \citep[e.g.,][]{Dole64,KaneHinkel13,Forgan2016,Popp2017}, past works have not considered how the long-term tidal evolution of the host binary stars can impact the orbital evolution and habitability of CBPs. In this paper, we perform simulations of a single planet orbiting two stars with a binary orbital period ($P_{orb}$) less than $\sim$10 days, that is, where tidal forces are expected to strongly impact the binary orbit \citep{ZahnBouchet89,Meibom2005,Fleming18}. We examine the orbital stability and long-term orbital oscillations of potentially habitable CBPs to determine the conditions that permit binary stars to host habitable CBPs, as well as to predict the properties and evolutionary histories of potentially habitable planets in those systems.

The orbits of binary stars span a wide range of both orbital periods and eccentricities \citep[\eg][]{Duquennoy1991,Raghaven10}, but nearly all binaries with orbital periods less than about 10 days are on circular orbits \citep{Meibom2005,Raghaven10}. This observation is likely due to orbital circularization caused by tidal forces that are strongest during the pre-main sequence phase \citep{ZahnBouchet89}. Low-mass stellar binaries with sufficiently short orbital periods, $P_{orb} \lsim 10$ days, experience tidal forces that can gravitationally deform the stellar shapes, creating tidal bulges that are maintained by friction within the stellar convective envelopes \citep{Mazeh2008,Zahn2008}. Under equilibrium tidal theories \citep[see][]{Darwin1880,FerrazMello08}, these tidal bulges induce torques that drive secular changes in the binary orbital and stellar rotational angular momenta. Stellar evolution further complicates the story as stars contract and develop a radiative core during the pre-main sequence \citep{Baraffe15}, changing each star's angular momentum evolution.  Thus, the binary's orbital semi-major axis \abin and eccentricity \ebin can change significantly over time due to both stellar evolution and tidal forces.

The coupled stellar--tidal evolution of binaries is complicated and can generate additional observable features beyond the circular orbits of short-period binaries. \citet{Fleming18} showed that for binary orbits with initial orbital periods less than about 8-10 days, the binary orbit will expand and become more eccentric as tidal torques transfer stellar rotational angular momentum into the binary orbit as the stars, in turn, contract along the pre-main sequence and tides drive the binary toward the tidally locked state. Following the earlier work of \citet{Verbunt1981}, \citet{Fleming18} showed that the binary orbit eventually decays due to the removal of angular momentum via magnetic braking after the stars' rotational frequencies are synchronized to the orbital frequency. \citet{Fleming18} demonstrated that, in general, larger initial orbital periods result in less expansion and a slower decay to circular orbits since these larger initial orbits host a larger initial reservoir of angular momentum. They found that the maxima in \abin and \ebin occurred about 2 Myr after the start of the \citet{Baraffe15} stellar models. \cite{Fleming18} dubbed their model the ``Stellar-Tidal Evolution Ejection of Planets'' (STEEP). We note that \cite{Zoppetti19} showed that outward-then-inward evolution of the binary orbit can occur even when one ignores the stellar evolution.

Some low-mass binary stars are known to host CBPs and the populations appear to have been sculpted by the process outlined in \citet{Fleming18}. For example, while more than ten transiting CBPs have been detected in the \kepler field, such as Kepler-34 and 35 \citep{Welsh12TransitingCP}, no CBPs have yet been discovered orbiting binary stars with $P_{orb} \lsim 7.5$ days, the orbital period of Kepler-47 \citep{Orosz12,Winn15,Windemuth2019}. This curious lack of CBPs may be due to small number statistics \citep{Windemuth20}, or it may be real and represent a signature of physical effects that significantly reduces the formation and/or long-term stability of these planets. \citet{Munoz2015}, \citet{Martin2015} and \citet{Hamers2016} suggested that a third star on a wide orbit could destabilize planets orbiting binary stars with orbital periods in the range 3 d $\lsim~P_{orb}~\lsim$ 7.5 d. More recent research by \cite{Fleming18} showed that the outward-then-inward orbital evolution described above can destabilize the innermost planet of a system for all binary periods less than 8 days. They also showed that this process can set off a chain reaction that ejects additional planets. Thus, the coupled stellar-tidal evolution of binary stars can dramatically affect CBP orbits and motivates additional investigations in the role of coupled stellar--tidal evolution of binary stars on the stability, evolution, and habitability of CBPs. 

The \citet{Fleming18} model only employed one tidal model (the ``Constant Phase Lag" equilibrium tidal model \citep[CPL;][]{FerrazMello08}), and did not include the change in stellar density profile as stars evolve on the pre-main sequence, a property often parameterized as the radius of gyration, $r_g$. \citet{Fleming2019} added this phenomenon in their study of \kepler binary stars and found it significantly impacts the angular momentum evolution of the binary. Thus, a reanalysis of the STEEP method with this ``dynamic \rg'' model and a quantitative comparison with the ``static \rg'' model would help determine if the STEEP process is still a viable explanation for the lack of CBPs orbiting short-period binaries.  Moreover, they did not explore the impact on the circumbinary habitable zone \citep[HZ;][]{Kasting93,Kopparapu13,Cukier19} or systems that are not engulfed by instability, that is to say, the binary orbit evolves, but not significantly enough to eject the planets. In such cases, the tidal evolution of the stars does not remove a planet, but slowly changes its orbit, which can still have a profound affect on the climates and habitability of planets \citep[\eg][]{Spiegel10,Forgan2016,Popp2017,HaqqMisra19,Deitrick18b}. Thus, a fuller range of models and system parameters must be explored to comprehensively understand the history and properties of CBPs.

The Kepler-47 system is a crucial benchmark for assessing the role of coupled stellar--tidal evolution. The binary orbit is $\sim 7.5$ days \citep{Orosz12}, and its inner planet, Kepler-47 b, orbits very close to the critical semi-major axis for stability, \acrit~\citep[see][]{Holman99,Winn2005}. This system is therefore in the transition region from tidally interacting to non-tidally interacting binaries, yet possesses 3 planets. Moreover, the Kepler-47 planetary system contains a gas giant (Kepler-47 c) orbiting within the HZ of the binary stars \citep{Orosz12}, and so may possess a habitable exomoon \citep{HellerBarnes13}. Thus, any theory of CBPs orbiting tidally interacting binaries must reproduce this system, preferably without any fine tuning of parameters.

Although Kepler-47 is the only known tidally interacting binary to host a planet, the formation of CBPs in short-period binaries is likely. For example, \cite{Alexander12} found that circumbinary disks form in a very similar process as single stars' disks for binary semi-major axes $\lessapprox 1$ au. Moreover, \citet{Bromley15} show that planetary formation around binary stars is similar to those that form around single stars. They also concluded that circumbinary disks and planets likely form in a similar manner as those hosted by single stars, implying that there should be a sizeable population of short-period CBPs forming during the early stages of binary star evolution, but none have yet been discovered.

To help resolve the outstanding issues, we performed new and more sophisticated simulations of the coupled stellar-tidal-orbital evolution of known and hypothetical systems consisting of two stars orbited by one planet and report the results here. In Section 2, we describe our models and the experimental setup of our simulations. In Section 3, we present results for a) the STEEP process with a second tidal model \citep[the ``Constant Time Lag" equilibrium tidal model, CTL;][]{Leconte10} and an evolving mass concentration, b) an extended exploration of the history of the Kepler-47 system, c) the evolution of the HZ and \acrit~for equal-mass binary stars, and d) a reanalysis of the stability of CBPs in the HZ of tidally interacting equal-mass binary stars. In Section 4, we discuss our results and conclude. We note that we have made all source code and plotting scripts publicly available, and each figure caption concludes with a link to a GitHub repository for reproducibility (electronic version only).


\section{Methods}

In this section we describe our model, implemented in the \vplanet software package \citep{Barnes_2020}, which includes validated models for stellar evolution, the equilibrium tidal evolution of binary stars, and the orbital evolution of a single CBP. In the following the subsections, we briefly review \vplanet (section 2.1), the circumbinary HZ in section 2.2, the circumbinary stability limit in section 2.3, and our initial conditions and assumptions in section 2.4. For thorough discussions, derivations, and validations of the numerical methods and physical models used herein, the reader is referred to \citet{Barnes13,Baraffe15,Fleming18,Fleming2019} and \citet{Barnes_2020}.

\subsection{\vplanet}
\vplanet is a software package designed to connect physical models that can be described by ordinary differential equations or explicit functions of time. \citep{Barnes_2020}. Each ``module'' is a physical process that can be selected at runtime. In our case, we focus on the time evolution of binary stars and their CBPs using the modules \stellar (stellar evolution), \eqtide (tidal evolution), and \binary (CBP orbital evolution), as described in the next section. During a simulation, each module operates simultaneously to compute time derivatives of each model parameter, advancing the system forward in time and permitting gigayear simulations of common astrophysical phenomena. \vplanet uses a fourth order Runge-Kutta scheme with dynamical timestepping that minimizes computational expense while still resolving all processes, regardless of timescale. A tuneable parameter provides the desired accuracy; here we assume convergence when a change of a factor of 10 in the timestep results in a change of less than $10^{-4}$ in the outcome. 

The \stellar module simulates the pre-main sequence and main sequence evolution of stars by interpolating the \citet{Baraffe15} radius, effective temperature, and radius of gyration ($r_g$, mass concentration) tracks. Angular momentum is conserved during the stellar evolution, so changes to these properties affect the rotational frequency (see \citet{Fleming18,Fleming2019} for a detailed description of our implementation of this effect). \stellar also includes several magnetic braking/stellar wind models that also affect the rotational evolution. We use the \citep{Matt15} model that is calibrated to \kepler observations of single stars to simulate magnetic braking for our low-mass binary stars. 

The \eqtide module employs the equilibrium tide model originally described in \cite{Darwin1880}, including both the CPL and CTL tidal incarnations \citep[see][]{FerrazMello08,Greenberg09,Leconte10}. These two models include ODEs for rotational frequency, obliquity, semi-major axis, and eccentricity for a two-body system \citep{FerrazMello08,Leconte10} that are appropriate for binary orbital eccentricities$\lsim 0.3$. 

The \binary module is an implementation of the approximate analytic model for the orbital evolution of a CBP derived by \citet{Leung2013} that is valid for small CBP orbital eccentricities and inclinations. It treats the planet as a massless particle whose motion is completely controlled by the evolving gravitational field of the stars. 

The \eqtide and \stellar modules were combined in \citet{Fleming2019} to reproduce the observed distribution of \kepler star rotation periods, including the population of weakly tidally interacting binaries with rotational periods of about 90\% the orbital period. These stars' rotation periods are in a quasi-equilibrium such that the change in rotational frequency due to contraction is about equal to the frequency damping due to the tidal torque. We use the \citet{Fleming2019} models here, with all output produced by \vplanet v1.1.

\subsection{The circumbinary habitable zone}
We estimate the circumbinary HZ limits given by \cite{Kopparapu13} that are defined by the effective stellar flux. The inner edge of the HZ is defined by the runaway greenhouse effect in which the oceans of an Earth-like planet are completely evaporated and lost to space. The outer HZ is defined by the maximum greenhouse effect, where CO$_2$ in an Earth-like atmosphere condenses, reducing the greenhouse effect along with the planet's surface temperature such that the planet remains permanently and globally ice-covered.

We restrict the investigation to equal-mass binary stars since the HZ model from \citet{Kopparapu13} focuses on HZs of a single spectral energy distribution (SED). While we could have employed the unequal stellar mass HZ models of \citet{Cukier19}, we only consider equal-mass stars to limit the parameter space and identify first order trends. We assume equal-mass binary stars emit effectively the same SED so we can safely employ the \citet{Kopparapu13} HZ limits by simply doubling the stellar luminosity term. Our model is functionally identical to that in \citep{Eggl18} The \citet{Kopparapu13} HZ limits for single and binary stars are implemented in \vplanet and therefore we can trivially track those limits as a binary system evolves.

We note that the CBP HZ limits vary as a function of the binary's orbital phase as the stars trace out their orbits \citep[e.g.,][]{Haghighipour2013}. In this work we are focused on the evolution of the binary and CBP orbits over billions of years, so we instead neglect those smaller variations by averaging the flux received by the CBP over the binary orbit. Moreover, climate models have found that CBP surface temperatures are generally insensitive to instellation variations that are typical for CBPs \citep{Popp2017}.

\subsection{The circumbinary stability limit}
Both \citet{Dvorak1986} and \citet{Holman99} characterized the circumbinary stability limit through numerical simulations and found that it lies at approximately 1.6 times the binary's semi-major axis, but is also a function of both the binary orbital eccentricity and mass ratio. \citet{Holman99} performed an extensive suite of simulations and developed an accurate analytic fit for this boundary, referred to by those authors as the critical semi-major axis that is given by the following relation
\begin{equation}
\begin{split} \label{eqn:crit_semi}
a_{crit} = & (1.60 + 5.1e_{bin} -2.22 e_{bin}^2 + 4.21 \mu_{bin} \\
& -4.27e \mu_{bin} -5.09 \mu_{bin}^2 + 4.61 e^2 \mu_{bin}^2) a_{bin} , \\
\end{split}
\end{equation}
where $a_{bin}$ is the binary orbital semi-major axis, $e_{bin}$ the binary orbital eccentricity, and $\mu_{bin} = m_2 / (m_1 + m_2)$ is the binary mass ratio (with $m_1$ the primary's mass and $m_2$ the secondary's mass. This value is calculated by \vplanet so we track its value in our simulations as each binary--planet system evolves.

\subsection{Initial conditions and assumptions}

With the exception of Kepler-47, we are interested in the evolution of a typical CBP system and therefore must define our parameter space and key model parameters. We limit our studies to stellar masses less than 1.1\Msun~ and ignore the post-main sequence evolution. We consider binaries with $P_{orb} \le 10$ days and $e_{bin} < 0.5$ to encompass the range of binaries for which tidal evolution is plausible and within the confines of the \eqtide module (larger eccentricities can induce effects that are not included). We will use initial rotation periods of 4 days as a fiducial value, but we do change this value in some cases, as noted below. The \citet{Baraffe15} stellar tracks are uncertain at early ages as the protostars are still coalescing out of the molecular cloud and likely harbor a circumbinary disk that can strongly impact the binary orbital evolution \citep{Fleming2017}. Following other investigations \citep{Fleming18,Fleming2019}, we set the initial age of the systems to be 1 Myr and assume that any planets have formed by this time. 

In \citet{Fleming18}, the role of an evolving mass distribution inside the star was not included even though it is a crucial component of accurate stellar evolution models \citep[\eg][]{Dotter08,Baraffe15}. Assuming that a star is initially a spherical distribution of gas, it eventually collapses and forms a dense H-fusing core, decreasing \rg~. For this study, we include the $r_g$ evolution from the \cite{Baraffe15} models as implemented by \citet{Fleming2019} in an update to \stellar.

Tidal parameters in stellar binaries are uncertain. Different investigations have found that tidal parameters can range over several orders of magnitude and still be consistent with observations \citep[\eg][]{Zahn89,Jackson09,Fleming2019}. In this study, we use both the CTL and CPL formulations of equilibrium tidal theory. Tidal effects in this model are parameterized by the Love number of degree 2, $k_2$, and the tidal time lag, $\tau$, for the CTL model or the ``tidal quality factor,'' $Q$, for the CPL model. The $k_2$ parameter is essentially a measure of the height of the tidal bulge and ranges from [0,1.5]. For our simulations, we assume $k_2 = 0.5$. Both $\tau$ and $Q$ are related to frictional forces inside the star that prevent the bulge from aligning perfectly with the centers of mass of the tidal perturber. The tidal $Q$ quantifies the phase offset between the tidal bulges and the line connecting the centers of mass of the tidally interacting bodies \citep{FerrazMello08}. Smaller values of $Q$ lead to more rapid evolution as tidal effects scale as $k_2/Q$ \citep{FerrazMello08}. Under the CTL model, $\tau$ quantifies the time lag between the passage of the tidal perturber and the tidal bulge \citep{Leconte10}. Larger values of $\tau$ result in faster evolution as the tidal effects scale as $k_2 \tau$. Therefore, there is a degeneracy between $k_2$ and $Q$ and $\tau$, so we opt to hold $k_2$ fixed and instead vary $Q$ or $\tau$ to quantify how the tidal dissipation impacts the evolution. 

We adopt $\tau = 0.1$ seconds \citep{Leconte10} for our fiducial value for our CTL model. For the CPL model, we use $Q = 10^6$ as our fiducial value as it tends to be consistent with observations \citep[\eg][]{Jackson09,Fleming2019}. For simulations that include a CBP, we keep its inclination $i$ $\lessapprox 0.25^\circ$ to avoid any issues with the approximations made by the \citet{Leung2013} derivation as implemented in the \binary module. We also ignore tidal effects on the CBP, as well as any torques on its rotation.

\section{Results}

In this Section, we examine the results of several suites of simulations that elucidated how the coupled stellar--tidal evolution of binary stars can impact the evolution of their planets. In Section 3.1, we consider the STEEP process with the CTL model and an evolving internal stellar mass distribution to build on the \citet{Fleming18} limited exploration. In Section 3.2, we perform detailed simulations of Kepler-47 to further understand how its planetary system avoided disruption and in Section 3.3 we consider the orbital evolution of a stable CBP orbiting a short-period stellar binary. Finally, in Section 3.4 we compare the circumbinary HZ to $a_{crit}$ for equal-mass binary stars to constrain where potentially habitable CBPs can exist.

\subsection{Revisiting STEEP}

\begin{figure}
	\includegraphics[width=\columnwidth]{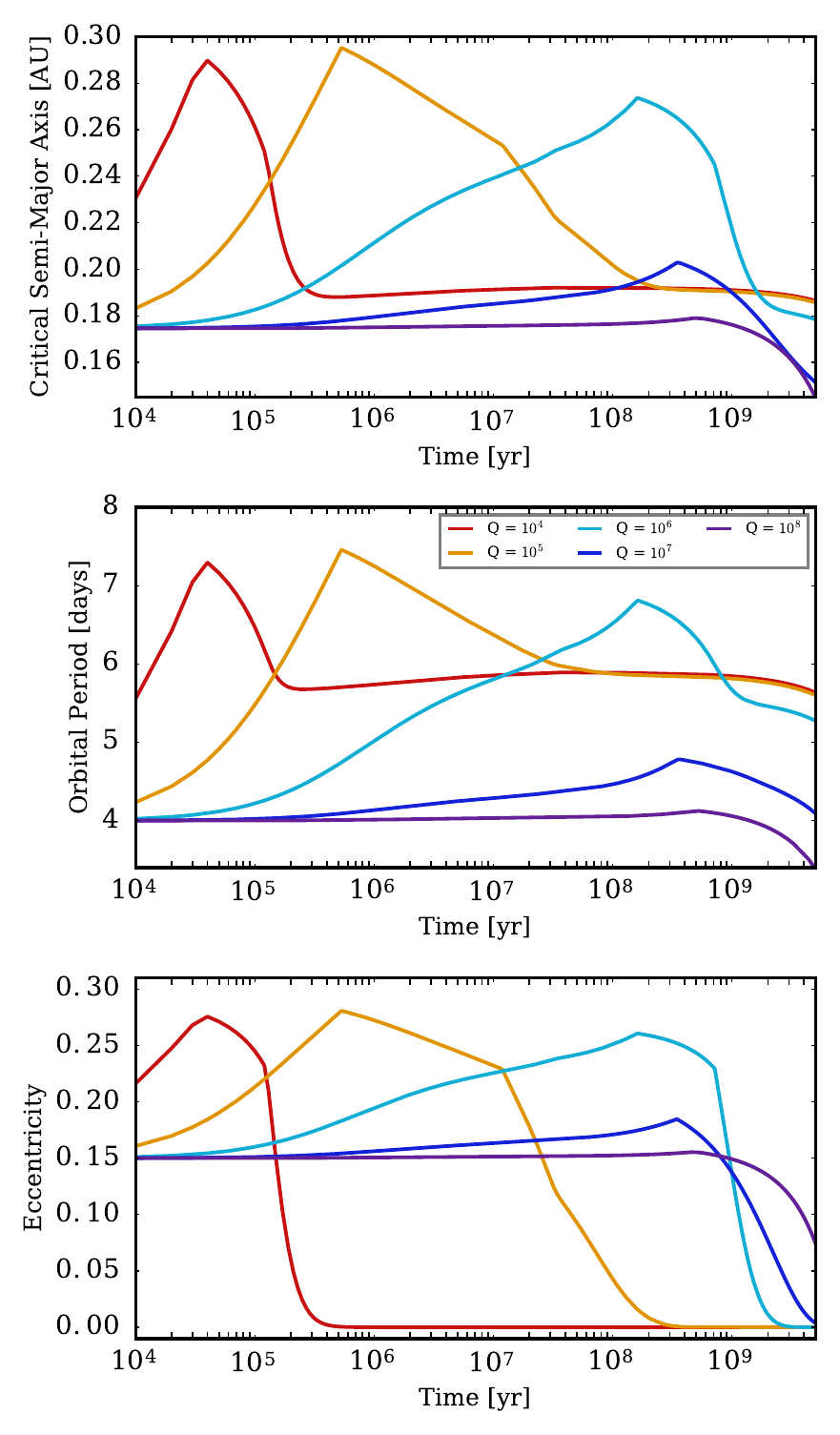}
   \caption{ Dynamic evolution of the star system from \citet{Fleming18}, influenced by different initial Q factors using the CPL model. Redder curves represent low initial Q values, while bluer curves represent higher Q values. We used the CPL model along with dynamic radius of gyration. {\it Top:} Binaries' dynamic stability limit time evolutions, \acrit. {\it Middle:} Orbital period of the secondary star. {\it Bottom:} Eccentricity of secondary star.  The approximate runtime was 15 minutes. \blue{\href{https://github.com/dglezg7/cbp_dynamic_stability/tree/master/Qfactors}{\link{github.com: cbp\_dynamic\_stability/Qfactors/}}}}
    \label{tidalq}
\end{figure}

We first simulate STEEP with a dynamic $r_g$ over a range of tidal $Q$s to assess how its inclusion modifies the results from \cite{Fleming18}. We limit our reanalysis to the equal-mass binary case (1M$_\odot$) shown in their Fig.~6. The top panel of Fig. \ref{tidalq} demonstrates that \acrit~ reaches its largest value when $Q = 10^5$. The $Q=10^5$ simulation achieves the highest \aratio~ ratio of roughly 1.69, compared to 1.1 for $Q=10^7$. This behavior is also seen in the middle panel, where once again the $Q = 10^5$ stars achieve the longest orbital period. Finally, the bottom panel shows the similar behavior, which shows how the stars reach a peak in eccentricity, followed by a rapid drop due to tidal circularization. We also note that as tidal $Q$ increases, these parameters reach their maximum values at later times: At $Q = 10^4$, the maximum \acrit~ occurs at 40,000 years while at $Q = 10^8$, the maximum \acrit~ arrives after about 500 Myr. For $10^4 \leq Q \leq 10^6$, the peaks shift mostly to later times, while for $10^6 \leq Q \leq 10^8$, the peaks occur at about the same time, but are significantly smaller. $Q = 10^6$ appears to be a transitional value in which the peak occurs relatively late, but is still relatively high.

\begin{figure}
	\includegraphics[width=\columnwidth]{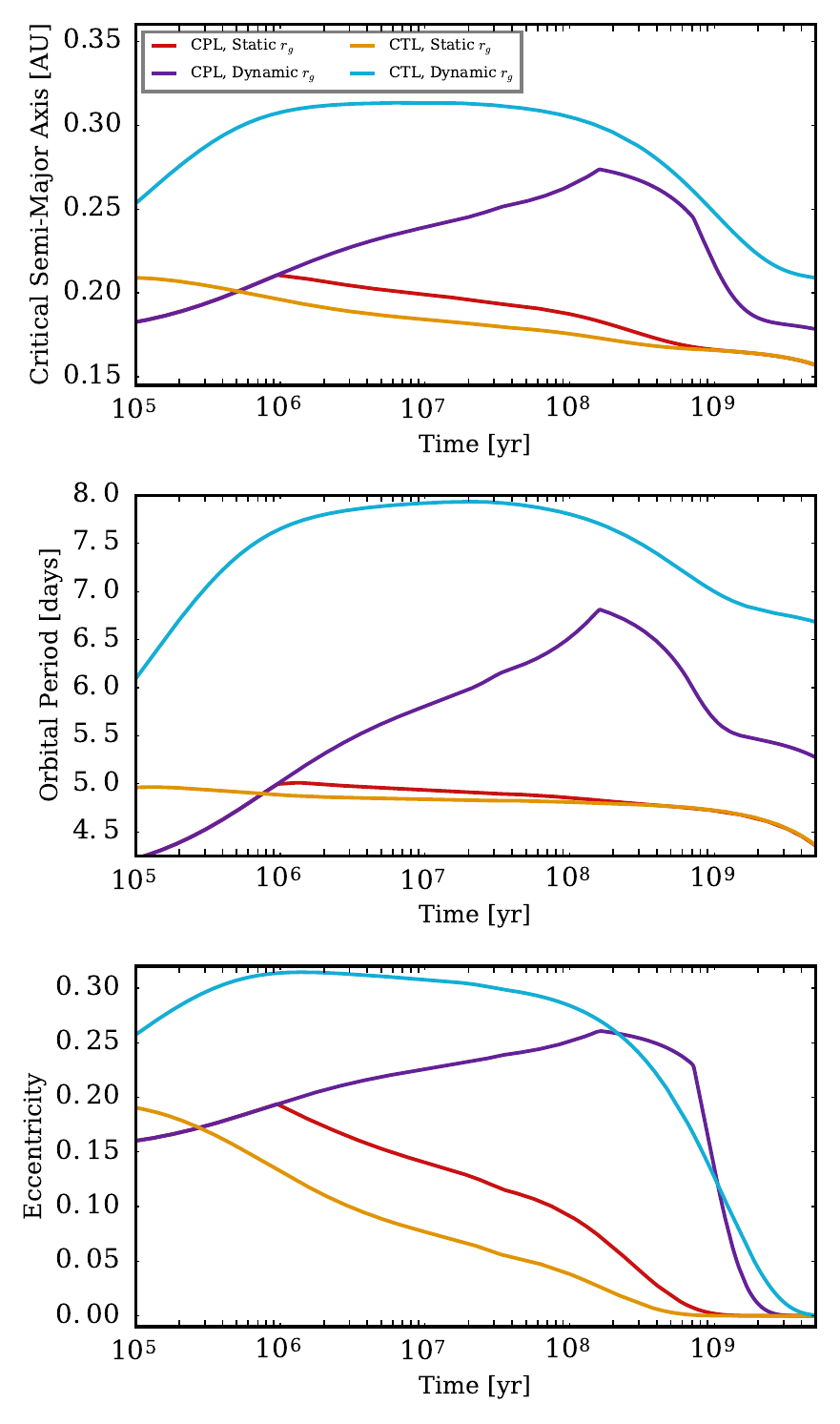}
   \caption{Long-term coupled stellar--tidal evolution of two solar-mass stars for various assumptions. Red curves represent stars that experience the CPL tidal model and constant radius of gyration $r_g$; purple curves are the CPL with evolving $r_g$; orange curves are the CTL tidal model with constant $r_g$; and pale blue curves are the CTL with evolving $r_g$. {\it Top:} Evolution of the stability limit \acrit. {\it Middle:} Orbital period of the binary star. {\it Bottom}: Binary's orbital eccentricity. The approximate runtime was 40 minutes in total. \blue{\href{https://github.com/dglezg7/cbp_dynamic_stability/tree/master/STEEP}{\link{github.com: cbp\_dynamic\_stability/STEEP/}}}}
    \label{steep}
\end{figure}

We next present results from four simulations designed to quantify the difference between the CPL and CTL models in the STEEP process. Our control is the fiducial example from  \cite{Fleming18}, which consists of two solar-mass stars with $Q=10^6$ with an orbital period of 4 days and  $e_{bin}$ = 0.15. The four simulations are the four permutations of our tidal models (CPL, CTL) and mass concentrations (dynamic $r_g$,  static $r_g$). The results are shown in Fig.~\ref{steep}. The red curves are the fiducial case examined in  \citet{Fleming18}. 

The top panel of Fig.~\ref{steep} shows \acrit~and we clearly see that the dynamic $r_g$ cases achieve much larger values than the static cases,. The difference between \acrit~growth between dynamic and static cases for the CPL and CTL models is about 25\% and 50\%, respectively. We find that larger initial \rg~produces larger \aratio ratios that peak at later times. This result suggests that the analyses and predictions in \citet{Fleming18} underestimated the impact of pre-main sequence coupled stellar-tidal evolution on CBPs of short-period binary stars because their model omitted stellar \rg~evolution. Hence, the STEEP process remains a viable explanation for the lack of CBP detections for short-period binaries.

This panel also shows that the two tidal models behave qualitatively differently. The CPL model has an abrupt and discontinuous peak, whereas the CTL model has a smoother peak that arrives much earlier. In many ways these features are consistent with the assumptions of the tidal models. The CPL model is a discrete function of tidal frequencies whereas the CTL model assumes a continuous distribution \citep{Leconte10}. As discussed in \citet{Fleming18}, the CPL peak occurs shortly after the stars become tidally locked. The CTL model's continuous tidal frequency dependence allows \aratio to decrease smoothly after the stars tidally lock. The maximum values of \aratio are fairly similar between the models, with the differences here likely due to our choices of $Q$ and $\tau$.

\subsection{The Kepler-47 system}

 \begin{figure*}
\centering
   \includegraphics[width=\textwidth]{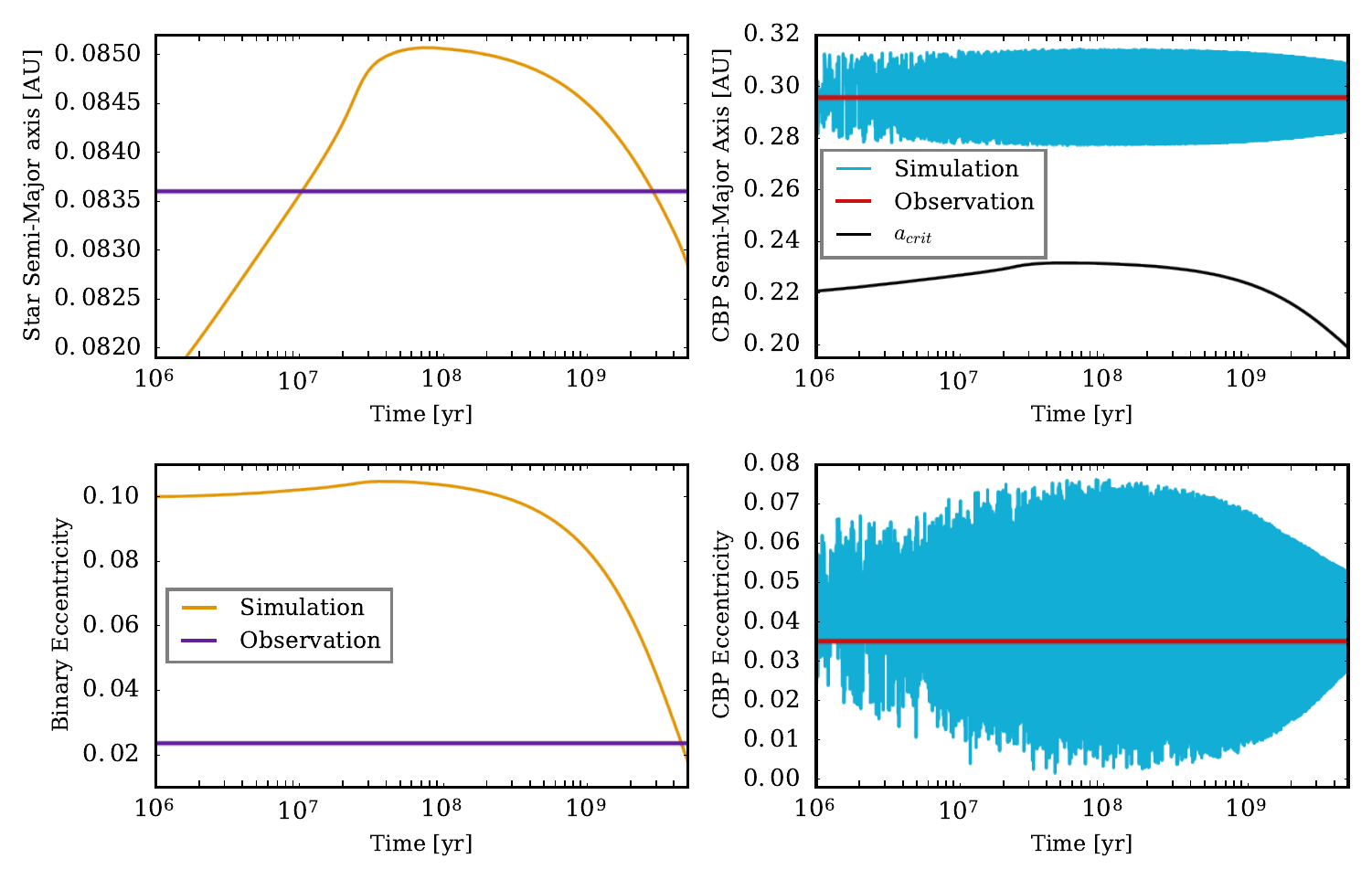}
     \caption{Orbital evolution of Kepler-47 B and Kepler-47 b. In each plot we compare the simulated values to their present day observed values. We can also see how the critical semi-major axis affects and raises the amplitude of the CBP's semi-major axis and eccentricity as the planet approaches the dynamic stability limit. The approximate runtime was 90 minutes. \blue{\href{https://github.com/dglezg7/cbp_dynamic_stability/tree/master/kepler47b}{\link{github.com: cbp\_dynamic\_stability/kepler47b/}}}}
     \label{k47b}
\end{figure*}

\begin{table}

\caption{Initial conditions for the Kepler-47 simulation.}              
\label{K47table}      
\centering                                      
\begin{tabular}{c c c}          
\hline\hline                        
Initial Value & K47 (A \& B) & K47b \\    
\hline                                   
    $P_{orb}$ [days] & 7.0 & 49.5 \\      
    $a$ [AU] & 0.080 & 0.2956 \\
    $e$ & 0.1 & 0.04 \\
    $i$ [deg] & 0 & 0.25 \\
    $P_{rot}$ [days] & 4.0 & --- \\
    $\tau$ [seconds] & 0.1 & --- \\
    $r_g$ & 0.27 & ---\\\\
\hline                                             

\end{tabular}
\end{table}

In light of the results above that reveal the importance of including $r_g$, we next reevaluate the Kepler-47 system because its inner planet borders the current critical semi-major axis. In Fig.~\ref{k47b}, we examine one plausible case in which Kepler-47 b stays safely away from the stability limit. In this simulation, we assumed that Kepler-47 b formed near its observed location, with initial conditions shown in Table 1. Thus, the STEEP model is still consistent with the Kepler-47 system.

\begin{figure}
	\includegraphics[width=\columnwidth]{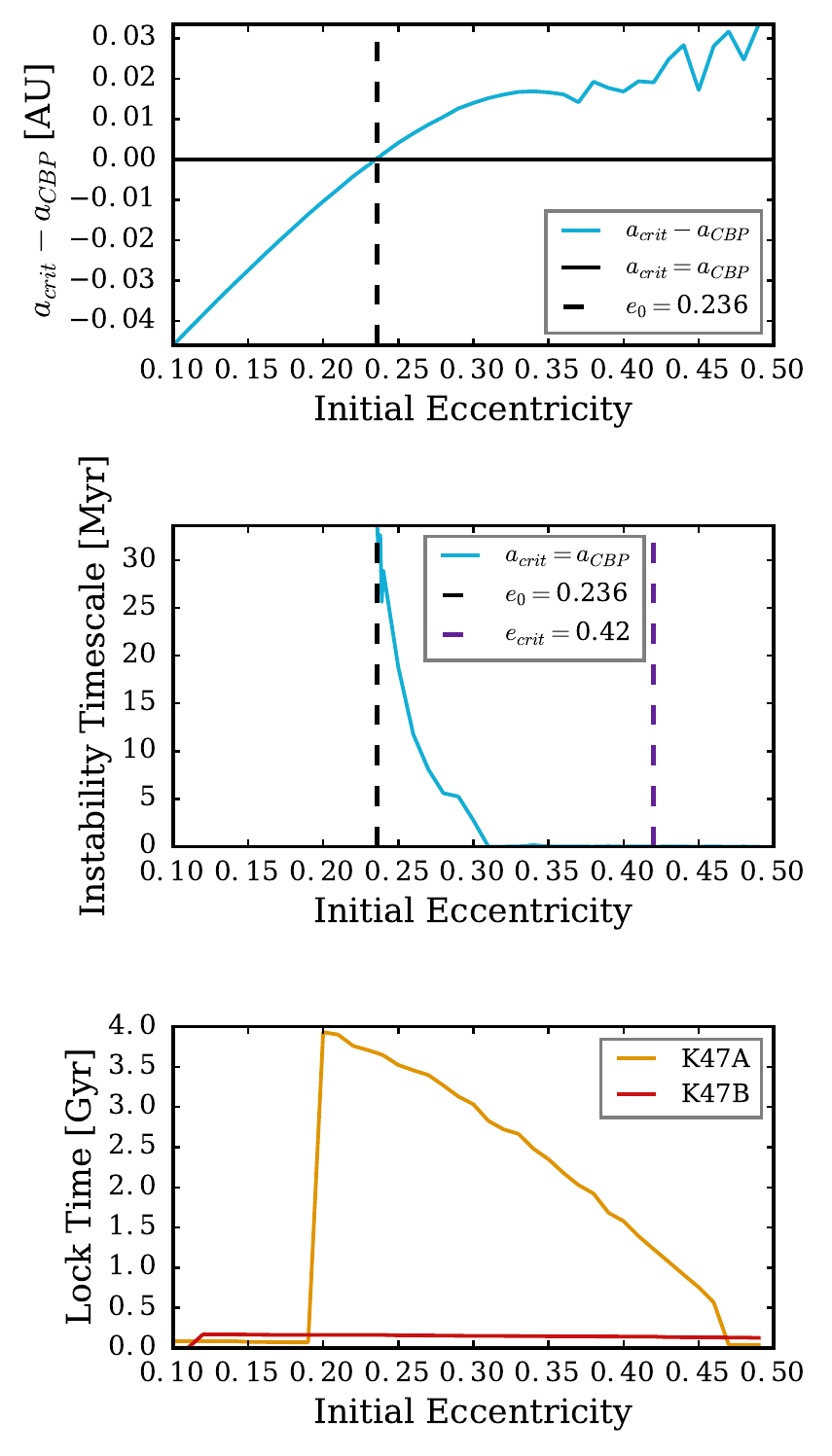}
   \caption{Dynamic stability and instability behavior of Kepler-47 b depending on the initial binary eccentricity. The top panel represents the maximum difference between \acrit~ and \acbp in its entire long-term evolution. If \acrit $-$ \acbp is less than zero, the dynamic stability limit never engulfs Kepler-47b. The middle panel shows the time in which \acrit~ crosses into \acbp (\acrit $=$ \acbp) during each simulation's history. The maximum instability time is 33.6 Myrs at \ebin $=$ 0.236 and drops as the initial \ebin increases. The dashed vertical lines represent transitional values of $e_{bin}$. The bottom panel shows when each of the binary stars tidally lock during each simulation. Tidal lock time for the secondary star, Kepler-47B, is constant compared to the initial \ebin, however the tidal lock time for the more massive primary star, Kepler-47A, jumps from 69.3 Myrs at \ebin $=$ 0.19 to 3.93 Gyrs at \ebin $=$ 0.20 and decreases to zero as the initial \ebin increases. The approximate runtime was 42 hours in total. \blue{\href{https://github.com/dglezg7/cbp_dynamic_stability/tree/master/K47_Eccentricities}{\link{github.com: cbp\_dynamic\_stability/K47\_Eccentricities}}}}
    \label{K47_Eccentricities}
\end{figure}

To further constrain the history of this system, we also performed simulations that varied the initial binary eccentricity. We first define the critical eccentricity $e_{crit}$ to be the initial eccentricity in which planet b's current semi-major axis is equal to \acrit. We used equation (\ref{eqn:crit_semi}) and found $e_{crit} = 0.42$. We then performed integrations over a range of initial $e_{bin}$ to identify values that could lead to the ejection of planet b. We find for our fiducial case that if the initial $e_{bin}$ was ${\lsim} 0.24$, then the instability zone never engulfed Kepler-47 b. For initial $e_{bin}$ between 0.24 and 0.42, $a_{crit}$ does reach b's semi-major axis. This result is visualized in Fig.~\ref{K47_Eccentricities} and further supports the STEEP model. We note that this example assumes initial rotation periods, so one should not conclude that the initial binary eccentricity must have been less than 0.24.

\subsection{Long-term CBP orbital evolution}

\begin{figure}
	\includegraphics[width=\columnwidth]{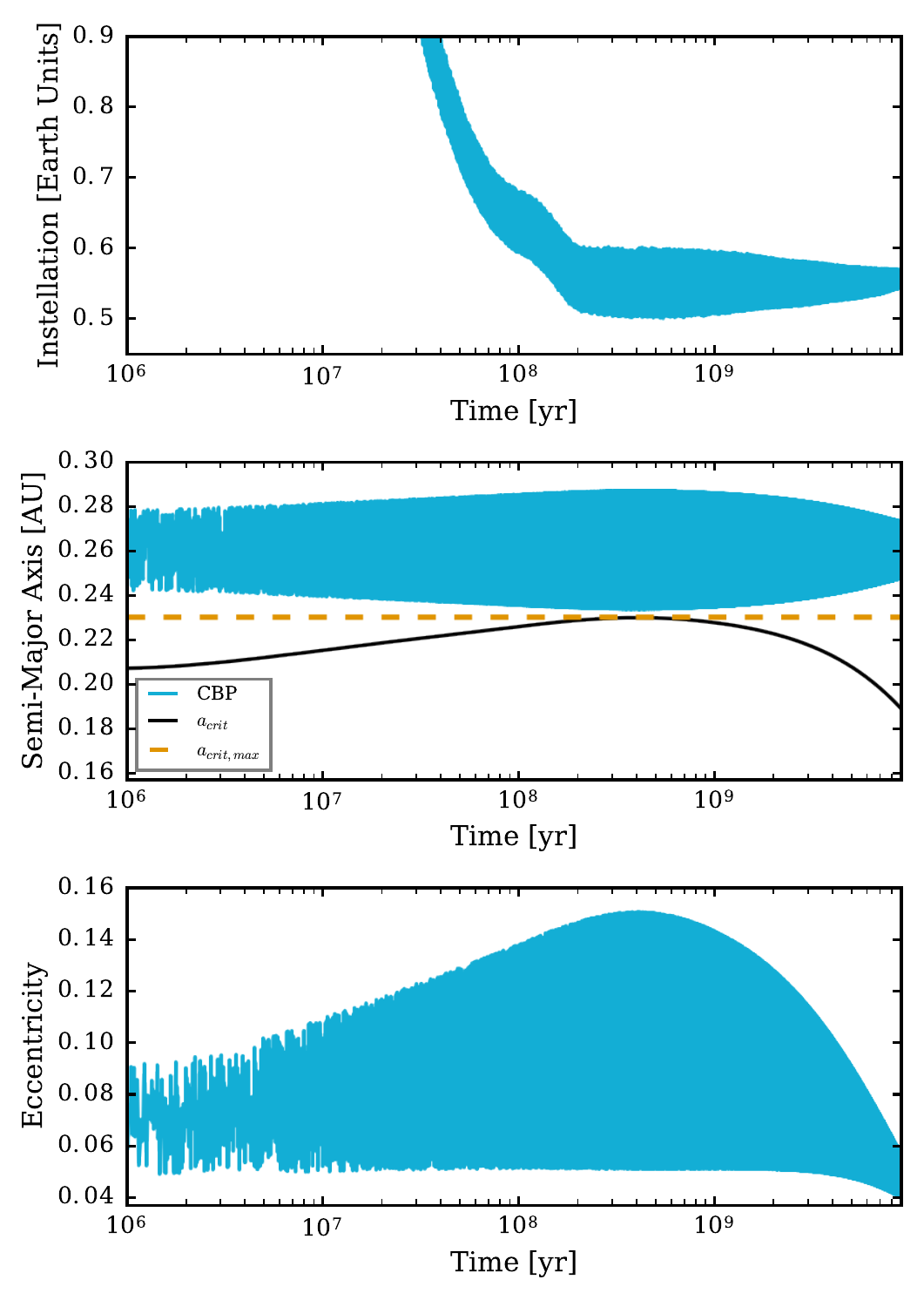}
   \caption{Instellation and the orbital evolution of a potentially habitable CBP with initial conditions listed in Table 2. The initial conditions of the binary stars and the CBP were adjusted so that the CBP can achieve large and variable amplitudes in its semi-major axis and eccentricity evolution due to gravitational perturbations from the central binary. The approximate runtime was 2 hours.  \blue{\href{https://github.com/dglezg7/cbp_dynamic_stability/tree/master/HabitableCBP}{\link{github.com: cbp\_dynamic\_stability/HabitableCBP/}}}}
    \label{habitablecbp}
\end{figure}

Next we consider the evolution of a potentially habitable planet orbiting a low-mass stellar binary to assess how the orbital evolution of the binary affects the CBP's long-term orbital evolution. To address this possibility, consider the system shown in Table 2, whose evolution is visualized in Fig.~\ref{habitablecbp}. In this case, the two stars are M dwarfs and the planet's semi-major axis remains beyond \acrit~ for the duration of the simulation. As the host stars are M dwarfs, their bolometric luminosity evolves considerably with time \citep[see \eg][]{Baraffe15}, so there is some danger the CBP may be rendered uninhabitable during its host stars' pre-main sequence phases \citep[\cf][]{LugerBarnes15}. Should the planet remain habitable after both stars begin fusing hydrogen, it will settle into a state in which its instellation fluctuates between 50--60\% of Earth's current insolation, as shown in the top panel of Fig.~\ref{habitablecbp}, which is well within the limits of the HZ \citep{Kopparapu13}.

The bottom two panels of Fig.~\ref{habitablecbp} show the CBP's orbital evolution. The semi-major axis oscillates with an amplitude of about 0.02 au, but the limits do evolve over time due to the binary star's orbit, which is not shown. The eccentricity evolution is considerably more dramatic, with an initial amplitude of $\sim 0.04$, growing to $\sim 0.1$, before damping to a small oscillation at late times. We note also that the ``floor'' of the eccentricity stays constant at about 0.05 until about 5 Gyr, at which point it also starts to decay. This system is probably evolving to the ``fixed point solution'' in which one secular frequency decays away and the major axes of the binary and the CBP circulate at constant frequency \citep{WuGoldreich02,Rodriguez11,vanLaerhoven14}. Thus, while hypothetical, this example demonstrates that the coupled stellar--tidal evolution of a binary star can significantly impact the orbital and instellation evolution of a potentially habitable CBP. 

In addition to this case, we performed numerous simulations that we do not show here. The example we do present was chosen because it demonstrated significant evolution -- most cases did not show such variations in the orbital evolution. In general we identified three traits that most correlated with strong variation in the CBP's dynamics: 1) nonidentical stellar masses, 2) a CBP's minimum semi-major axis that is close to $a_{crit,max}$, and 3) a roughly null initial CBP free eccentricity (\efree $\lsim 0.05$, see \citet{Leung2013}).

\begin{table}
\caption{Initial conditions for the system shown in Fig.~\ref{habitablecbp}}              
\label{habitablecbptable}      
\centering                                      
\begin{tabular}{c c c}          
\hline\hline                        
Initial Value & Star (A \& B) & CBP \\    
\hline                                   
    $P_{orb}$ [days] & 7.5 & 62.5 \\      
    $a$ [AU] & 0.063 & 0.26 \\
    $e$ ($e_{free}$ for the CBP) & 0.3 & 0.0 \\
    \Msun ($M_{\oplus}$ for CBP) & 0.5 \& 0.1 & 2.98 \\
    $P_{rot}$ [days] & 4.0 & --- \\
    $\tau$ [seconds] & 0.1 & --- \\
    
\hline                                             
\end{tabular}
\end{table}

\subsection{Habitable zone stability for equal-mass binaries}

 \begin{figure*}
\centering
   \includegraphics[width=\textwidth]{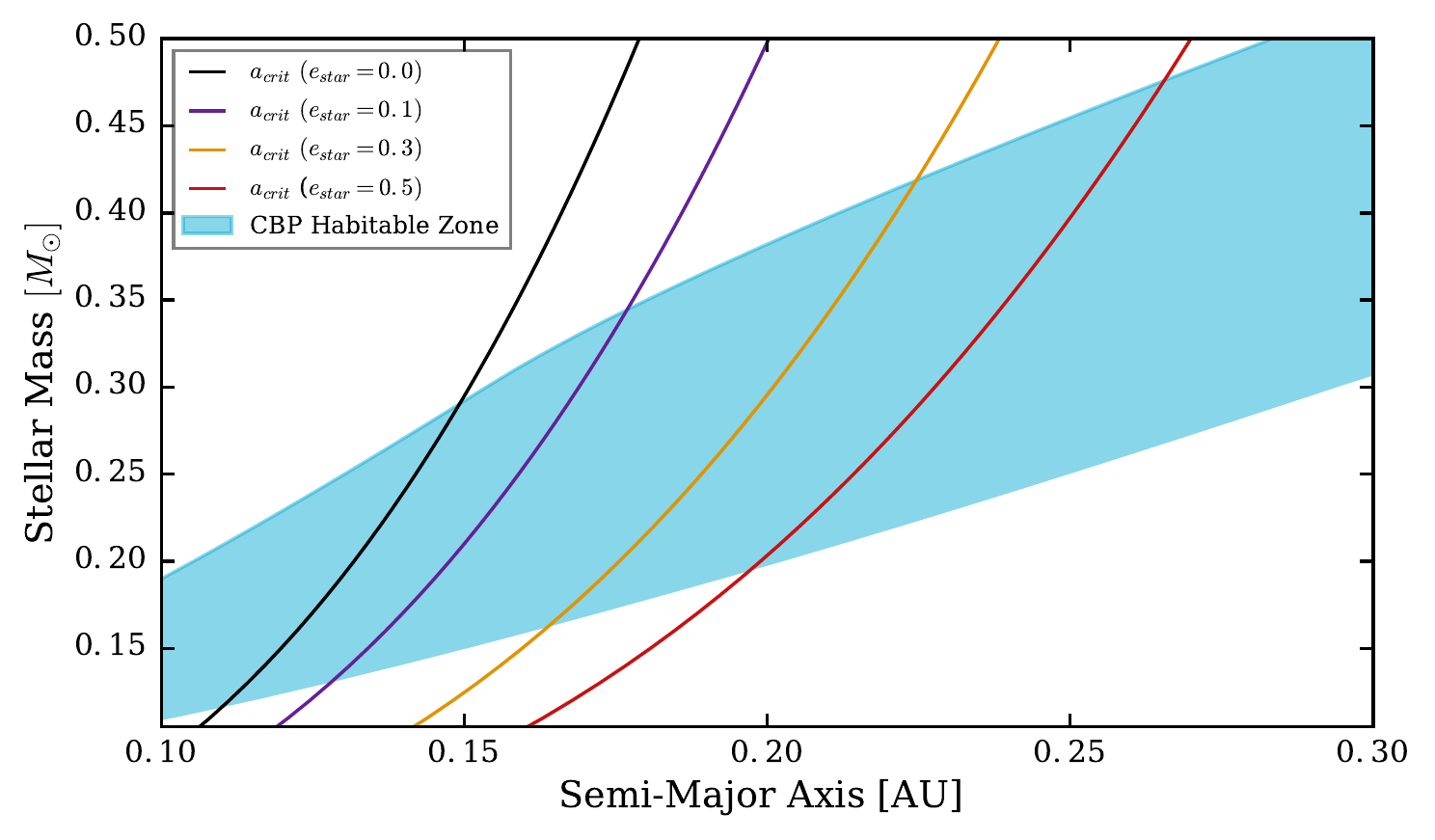}
     \caption{Comparison of the equal-mass binary HZ to \acrit. The curves correspond to different choices of the binary's orbital eccentricity, up to 0.5, as shown in the legend. The blue shaded region is the HZ from \citet{Kopparapu13}, but with the stellar luminosity doubled to account for the two stars. The approximate runtime was 15 seconds. \blue{\href{https://github.com/dglezg7/cbp_dynamic_stability/tree/master/HZ_CBP}{\link{github.com: cbp\_dynamic\_stability/HZ\_CBP/}}}}
     \label{hzcbp}
\end{figure*}

\begin{figure}
	\includegraphics[width=\columnwidth]{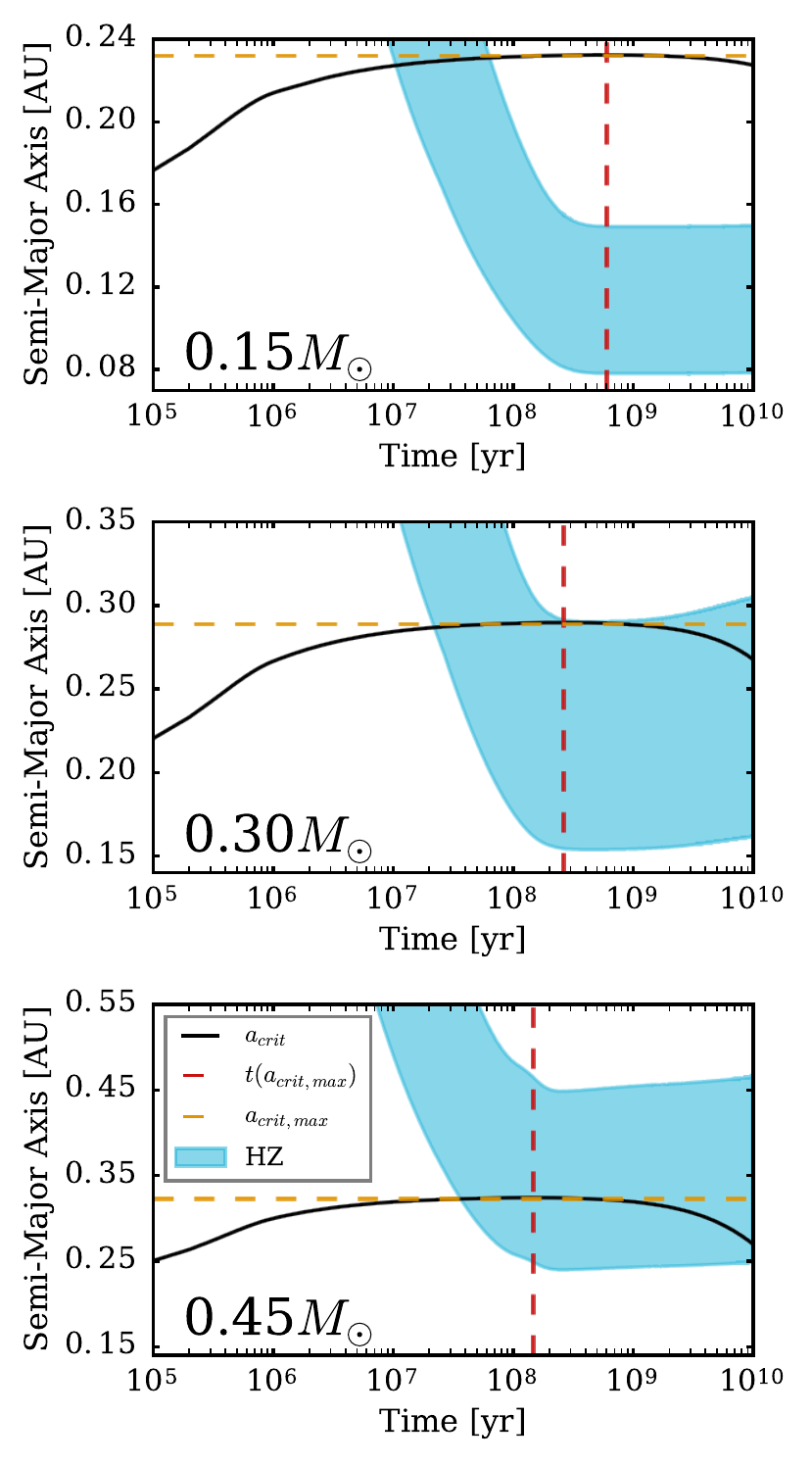}
   \caption{Evolution of \acrit~and the HZ for 3 hypothetical equal-mass binaries. The blue shaded region is the HZ; the black curves are the values of \acrit; the dashed horizontal orange lines are the maximum value of \acrit; and the dashed vertical red lines are the time that \acrit~reached its maximum value. {\it Top:} Stellar masses are 0.15\Msun. {\it Middle}: Stellar masses are 0.3\Msun. {\it Bottom}: Stellar masses are 0.45\Msun. The approximate runtime was 6 minutes in total. \blue{\href{https://github.com/dglezg7/cbp_dynamic_stability/tree/master/HZ_Evolution}{\link{github.com: cbp\_dynamic\_stability/HZ\_Evolution/}}}}
    \label{hzevolution}%
\end{figure}

Finally, we reconsider the stability of the HZ of equal-mass stellar binaries for which the CBP orbit is approximately coplanar with the binary orbit. We first overlay the location of \acrit~ with the HZ of \citet{Kopparapu13} in Fig.~\ref{hzcbp}. We assume the initial orbital period is 7.5 days and consider $0 \le e_{bin} \le 0.5$. We note that no stellar and/or tidal evolution is included in this figure, so it represents a conservative treatment of HZ stability since the stellar--tidal evolution almost always results in an expansion of \acrit~\citep{Fleming18}. We note for binaries with masses $\lsim 0.12$\Msun~that the entire HZ is unstable for all values of $e_{bin}$. On the other hand, we find that for stellar masses larger than 0.47\Msun, the entire HZ is stable for $e_{bin} \le 0.5$. These threshold stellar masses are listed in Table 3 as a function of initial $e_{bin}$. The ``Unstable HZ'' column refers to the maximum stellar mass for which the entire HZ is unstable; ``Stable HZ'' refers to the minimum stellar mass for which the entire HZ is stable.

Including the stellar--tidal evolution will generally push the stability limits out another 10--50\%, depending on the initial eccentricity and stellar rotation periods, as demonstrated above. Rather than simulate this entire parameter space, in Fig.~\ref{hzevolution} we show three example evolutions, in which $e_{bin}$ is initially 0.3 and the rotation periods are initially 4 days. We note that the HZ moves inward as the stars contract at approximately constant temperature, which causes their luminosities to decrease. \citep{Hayashi61,Baraffe15}. Should any particular binary star become of interest, one can easily modify the script referenced in the caption to Fig.~\ref{hzevolution} to estimate the evolution and stability of the binary's HZ. 

\begin{table}
\caption{Stability limits of the HZ for equal-mass binaries}              
\label{limitshzcbp}      
\centering                                      
\begin{tabular}{c c c}          
\hline\hline                        
$e_{bin,0}$ & Unstable HZ (\Msun) & Stable HZ (\Msun) \\    
\hline                                   
    0.0 & 0.120 & 0.280 \\      
    0.1 & 0.135 & 0.345 \\
    0.3 & 0.160 & 0.420 \\
    0.5 & 0.180 & 0.470 \\
    
\hline                                             
\end{tabular}
\end{table}

\section{Discussion}

We have explored the time evolution of CBPs orbiting low-mass short-period binary stars to further explore how tidal effects and stellar evolution impact the habitability of CBPs. We first revisited the STEEP process with the CTL tidal model and included an evolving mass concentration and discovered that they predict a greater expansion of \acrit~ than in \citet{Fleming18} found, suggesting their results understated the threat to CBPs posed by coupled stellar-tidal evolution.  We reexamined the transitional stellar binary Kepler-47 \citep{Orosz12} and found that its planetary system's stability likely required the initial binary eccentricity to be below 0.2. We presented an example of a stable and potentially habitable CBP orbiting a short-period binary to show that instellation can change dramatically over time in at least some cases due to the tidally driven orbital evolution of the central binary. Finally, we reanalyzed the stability of the equal-mass circumbinary HZ and found that CBPs may be ejected from the HZ for stellar masses $<0.5$\Msun. Taken together, these results clarify where habitable CBPs can exist and how they might evolve around short-period stellar binaries.

By considering alternative tidal and structural models of stars, we have improved and clarified the model presented in \citet{Fleming18}. That theory remains the only one that explains the non-detection of CBPs orbiting any binary star with an orbital period less than 7.5 days without a nearby tertiary companion (aside from small number statistics). We find that the \citet{Fleming18} model is likely to underestimate the effectiveness of the STEEP process. For triple systems, the timescale for the orbital evolution to eject the planet can be arbitrarily long, whereas the STEEP process occurs shortly after formation. 

Furthermore, we found that when including the $r_g$ evolution, the peak in \acrit~ not only extends to a larger semi-major axis, it occurs $~\sim$100 Myr after the stars formed. \citet{Fleming18} found that \amax tended to occur at around $10^6$ years, or during the tail end of the circumprimary disk's lifetime, see red curve in Fig.~\ref{steep}. Our updated model now predicts that the peak occurs well after planet formation. We therefore conclude that the coupled stellar--tidal evolution of binaries stars is the likely mechanism that cleared out these planets shortly after they formed, thereby explaining the lack of observed CBPs in the \kepler field for all $P_{orb} \le 7.5$ days.

This conclusion is further supported by our more rigorous analysis of the Kepler-47 system. Although our improvements to the \citet{Fleming18} model put Kepler-47b at more risk for ejection, we found a large region of parameter space for which planet b remains safely beyond \acrit~ for Gyrs. This parameter space was narrow, and included the initial rotation periods and tidal $Q$s or $\tau$s, but the values we used are broadly consistent with star formation and tidal models \citep[see][for a review]{Fleming2019}. Future work could identify the boundary in parameter space for Kepler-47b to never cross \acrit~ to derive constraints on the tidal dissipation in the host stars. Kepler-47 remains a valuable laboratory for binary star models, tidal effects, and CBPs. 

Our coupled stellar-tidal model improved upon that in \cite{Fleming18}, but more effects should be considered in the future. We did not consider stellar obliquity, which could affect the magnitude of STEEP and the timing of the maximum value of \acrit~could also be different. Future research should explore this parameter's role on the CBP evolution and habitability. Dynamical tide models consider friction due to flows often predict more accurate results \citep[\eg][]{Ogilvie07,Bolmont16}, and they should be applied to this problem.  Future research could also explore how the tidal deformation of stars impacts stellar evolution.

For habitable planets orbiting a short-period binary, a slowly evolving climate is possible. In addition to any changes due solely to stellar evolution, the instellation cycles, in other words, Milankovitch cycles, can change in both frequency and amplitude. We found one case in which the instellation amplitude changes from 20\% to 10\% over several billion years. Should a potentially habitable CBP be discovered, simulations like those depicted in Fig.~\ref{hzevolution} could be performed to understand its history, and therefore its likelihood to be habitable. 

In some cases the circumbinary HZ may be unstable and surveys for habitable worlds should avoid them. While this point has been discussed before \citep[\eg][]{KaneHinkel13}, we have shown quantitatively where the stability limit overlaps with the HZ, at least for a limited parameter space. We find that habitable planets cannot exist around a binary star whose component masses are less than 0.12\Msun. For stellar masses up to 0.28\Msun~ at least some of the HZ is unstable. For a typical binary eccentricity of 0.3, those values increase to 0.15\Msun~ and 0.43\Msun. When factoring in stellar--tidal evolution, these limits will increase by 10--50\% depending on the initial distribution of angular momentum. Thus, we recommend that surveys for potentially habitable CBPs focus on systems in which the component masses are larger than at least 0.3\Msun, if the binary's orbital period is also less than 8 days. 

Future research could identify the stable HZ for nonidentical-mass binary stars similar to the analysis in Fig.~\ref{hzcbp} and further constrain where habitable CBPs may be stable. Although we did not perform that parameter sweep, some general trends emerged from our investigations that are worth noting. We found that the evolution of \acrit~ for nonidentical binaries can achieve a higher \amax when compared to their identical mass analogs. The closer a CBP is to \acrit, the higher the amplitudes in \acbp and \ecbp become, therefore the change in their amplitudes can vary more significantly, too. Additional research could also explore the role of inclination, which has can cause deviations from the \cite{Holman99} stability limit \citep{PilatLohinger03,Quarles20} and can significantly alter CBP orbital evolution, presumably affecting STEEP and habitability.

Our orbital models are approximate and therefore results at high eccentricity should be treated as preliminary. The gigayear evolution of a CBP system is very challenging for an N-body integrator as the binary orbital period is so short compared to the lifetime of the system. Furthermore, the timescales for the orbital motion are orders of magnitude smaller than the tidal timescales such that properly treating their combined evolution can result in a simulation that will take years to complete. Nonetheless, a first principles approach would provide better insight and confirmation of the results presented here. Switching to an N-Body method would also eliminate the uncertainty in Eq.~(\ref{eqn:crit_semi}), which is a fit to a large number of N-Body simulations \citep{Holman99}. 

We focused our habitability analysis on orbital evolution, but many other factors might be important and could be included in future work. CBP rotation rates may reach a stationary solution analogous to tidal locking \citep{Zoppetti19,Zoppetti_2020}, which could significantly affect climates \citep[see e.g.,][]{Pierrehumbert11,Yang13,DelGenio19}. The planets of M dwarf stars, such as those in Figs. \ref{hzcbp}--\ref{hzevolution}, may spend significant time interior the HZ where XUV photons can desiccate the planet \citep{LugerBarnes15}. Colliding winds could exacerbate atmospheric loss on planets orbiting binary stars \citep{Johnstone_2015}, although \citet{Zuluaga_2016} showed that magnetic field could shield some planets. As with single stars, the habitability of CBPs is complicated and depends on many factors, not all of which could be included in this study \citep[c.f.][]{MeadowsBarnes18}.

The detection of a potentially habitable planet around a binary star would be a landmark achievement. As transit surveys are biased toward detecting planets near \acrit~in orbit around short-period binaries, the scientific community must understand how the coupled stellar and tidal effects in binary stars impacts the detectability and habitability of such worlds. We have confirmed that habitable CBPs of short-period binary stars must run a gauntlet of dynamical effects that could trigger ejections, dramatic climate change, and possibly sterilization. The results presented here can help guide astronomers toward the discovery of a habitable CBP, as well as provide clues to the history of such fascinating worlds.

\vspace{0.5cm}

\noindent DPF was supported by NASA Headquarters under the NASA Earth and Space Science Fellowship Program - Grant 80NSSC17K0482. RB's contribution was supported by NASA Virtual Planetary Laboratory Team through grant number 80NSSC18K0829. We thank Siegfried Eggl for valuable input. This work also benefited from participation in the NASA Nexus for Exoplanet Systems Science (NExSS) research coordination network.

\bibliographystyle{aa} 
\bibliography{vplanet.bib} 
\end{document}